\documentclass{JHEP3}

\usepackage{amssymb}
\usepackage{amsmath}

\usepackage{graphicx}
\usepackage{graphics}
\usepackage{epsfig}

\usepackage{multirow}

\title{Growth and dissolution of spherical density enhancements in
  SCDEW cosmologies. }

\author{ Silvio A. Bonometto\\ INAF, Trieste Observatory \& Trieste
  University, Physics Dep., Astronomy Unit \\ Via Tiepolo 11, 34143
  Trieste, Italy}

\author{ Roberto Mainini\\ Physics Department G.~Occhialini,
  Milano--Bicocca University\\
Piazza della Scienza 3, 20126 Milano, Italy }

\abstract{Strongly Coupled Dark Energy plus Warm dark matter (SCDEW)
  cosmologies are based on the finding of a conformally invariant (CI)
  attractor solution during the early radiative expansion, requiring
  then the stationary presence of $\sim 1\, \%$ of coupled--DM and DE,
  since inflationary reheating. In these models, coupled--DM
  fluctuations, even in the early radiative expansion, grow up to
  non--linearity, as shown in a previous associated paper.  Such early
  non--linear stages are modelized here through the evolution of a
  top--hat density enhancement. As expected, its radius $R$ increases
  up to a maximum and then starts to decrease. Virial balance is
  reached when the coupled--DM density contrast is just 25--26 and DM
  density enhancement is $\cal O$$(10\, \%)$ of total density.
Moreover, we find that this is not an equilibrium configuration as, afterwards, 
coupling causes DM particle velocities to increase, so that the fluctuation 
gradually dissolves.
We estimate the duration of the whole
  process, from horizon crossing to dissolution, and find
  $z_{horizon}/z_{erasing} \sim 3 \times 10^4$. Therefore, only
  fluctuations entering the horizon at $z \lesssim 10^9$--$10^{10}$ are
  able to accrete WDM with mass $\sim 100\, $eV --as soon as it
  becomes non--relativistic-- so avoiding full disruption.
  Accordingly, SCDEW cosmologies, whose WDM has mass $\sim 100\, $eV,
  can preserve primeval fluctuations down to stellar mass scale.  }

\begin{document}

\section{Introduction}
LCDM cosmologies are highly performing {\it effective} models. What be
the physics behind the LCDM paradigm, this is the question. A set of
options descend from assuming General Relativity violation, at large
scales or low densities. But listing these and other options (see
e.g. \cite{amenti}) is out of our scopes, as here we focus on a
peculiar variant of a specific option, that Dark Energy (DE) is a
scalar field $\Phi:$ when self--interacting, infact, a scalar field
can exhibit a negative pressure approaching its energy density
($|p_\Phi| \sim |\rho_\Phi|$) \cite{DE}. Within the frame of these
models we then treat a specific question concerning SCDEW (Strongly
Coupled Dark Energy plus Warm dark matter) cosmologies
\cite{previous}.

These cosmologies, widely discussed also in the previous associated
paper (hereafter BMM) \cite{bmm}, are however quite a peculiar branch
of scalar field models, being based on a conformally invariant (CI)
attractor solution of background evolution equations, holding all
through radiative eras, and allowing then for significant $\Phi$ and
Dark Matter (DM) densities. Let us recall soon that such DM is coupled
to DE and distinct from warm--DM, although viable models, discussed in
BMM, allow DM components to share several features, as Higgs' masses
$m_w$ and $\tilde \mu \equiv m_c$ (for WDM and coupled DM,
respectively) and primeval densities.

The focus of this paper is then on the early evolution of spherical
density enhancement in SCDEW cosmologies. In fact, besides of being
peculiar for the behavior of background components, they also exhibit
specific features in fluctuation evolution and, in this paper, we show
that coupled--DM fluctuations grow, indipendently of other components,
 and approach non--linearity  well before all of them. By
  adopting a spherical top--hat model, we then follow them in the
  non--linear stages, until the virialization condition is fulfilled,
  so that the sphere is should stabilize at a given radius $R$ and
  density contrast $\Delta$. Somehow unexpectedly, however, such
  virial equilibrium condition is not permanent and the sphere seems
  doomed to 
total dissipation; all that occurs through radiative eras  and
  will be probably end up as soon as other components are able to take
  part to the spherical growth.

 More in detail, 
here we shall quantitatively follow the evolution of coupled--DM
fluctuations until their (temporary) virialization (phase I) ,
  also testing how results depend on the redshift when the horizon
  reaches the fluctuation size, and the amplitude the fluctuation has
  then.  The dependence on the model parameters will be also partially
  explored.
What is expected to happen later (phase II) is harder to explore
  analytically, and our aim is just to give an order of magnitude for
  the time taken by dissipation.

Our final aim amounts
to approach a determination of the low--mass transfer function in
SCDEW cosmologies, over scales that
linear algorithms are unable to treat. 
 In particular we aim at constraining
the minimal scale for fluctuation survival, in SCDEW models.
According to our approach, such scale lays
in the large stellar mass range. Let us outline that it should be so
in spite of DM particles sharing a (Higgs) mass $\cal O$$(100\, $eV).

From a quantitative side it is then worth recalling that the early
intensity of DM--$\Phi$ coupling, in SCDEW models, is fixed by an
interaction parameter
\begin{equation}
\beta = b \sqrt{16\pi/3}
\label{beta}
\end{equation}
expected to be $\cal O$$(10)$. The early density parameters,
\begin{equation}
\Omega_c= \Omega_\Phi/2= 1/2\beta^2~,
\label{omegas}
\end{equation}
for coupled DM and $\Phi$, keep then a constant value through
  radiative eras, as both components expand $\propto a^{-4}$, just as
radiative components.

 Let us then outline soon that we shall use the background metric
\begin{equation}
ds^2 = a^2(\tau) (d\tau^2 - d\lambda^2)
\end{equation}
$\tau$ being the conformal time and $d\lambda$ the line element.

The primeval CI expansion is then broken by the acquisition of the
tiny Higgs' masses at the electroweak (EW) scale. WDM and the spinor
field $\psi$ yielding coupled DM, in particular, are supposed to
acquire a mass $\cal O$$(100\, $eV), intermediate between light quark,
electrons, and neutrinos. The effective mass of the $\psi$ field,
below the Higgs' scale, then reads
\begin{equation}
\label{meff1}
m_{eff} = g_h m_p \exp[-(b/m_p)(\Phi-\Phi_p)] + \tilde \mu~.
\end{equation}
Here $b$ coincides with the coupling in eq.~(\ref{beta}), $\Phi_p$ is
the value of the scalar field extrapolated to the Planck time
according to the CI solution (even though unlikely to hold so early),
$m_p$ is the Planck mass.
 During the CI expansion, $m_{eff}$ is given just by the first
  term of the expression (\ref{meff1}), while $\Phi-\Phi_p =
  \tau_p/\tau~.$ Accordingly, in such era
\begin{equation}
m_{eff} \propto \tau^{-1}~,
\label{propto}
\end{equation} 
such behavior being gradually violated when the term $\tilde \mu$
acquires relevance. Let us however add that the value of $g_h$ enters
quantitative results only through the ratio
\begin{equation}
{\cal R} = {\tilde \mu /( g_h m_p)}
\label{calR}
\end{equation} 
taken here as effective parameter, so that we can fix $g_h =
  2\pi$ as in BMM. 

It is also worth defining $C = b/m_p$ and outline that the appearence
of a Higgs' mass bears another consequence, a progressive weakening of
the effective DM--$\Phi$ coupling; infact, as explained in detail in
BMM,
\begin{equation}
C_{eff} = {C \over 1 + {\cal R} \exp[C(\Phi-\Phi_p)]}
\label{beteff}
\end{equation}
so that, as soon as the dynamical (logarithmic) scalar field growth
makes $\Phi \sim \Phi_p - \log{\cal R}/C$, the effective coupling
weakens, and we expect $\beta_{eff}=\cal O$$(0.1)$ at the present
time.
\begin{figure}[t!]
\begin{center}
\vskip -1.5truecm
\includegraphics[height=10.cm,angle=0]{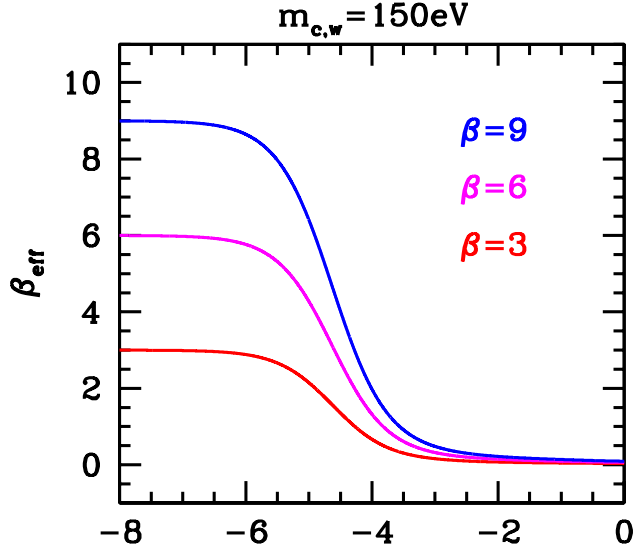}
\end{center}
\vskip -1.5truecm
\caption{Scale dependence of the effective coupling
  in SCDEW cosmologies with $m_{c,w} = 150\, $eV for various couplings
  $\beta$.  }
\label{betaeff}
\end{figure}

Accordingly, at $\tau < \tau_H$ ($T>T_H \sim 200\, $GeV, $z > z_H \sim
8 \times 10^{14}$), a full CI holds (the suffix $_H$ refers to Higgs'
mass acquisition at the EW scale).

Later on, at $\tau > \tau_H$ ($T<T_H \sim 200\, $GeV, $z < z_H \sim 8
\times 10^{14}$), CI is violated; however, being $\tilde \mu \ll T_H$,
a long period of {\it effective CI expansion} still occurs. Figure
\ref{betaeff} shows the gradual end of such effective CI expansion,
occurring quite late even for a fairly large value of $\tilde \mu
\equiv m_{c,w}$,
for several $\beta$ values. These behaviors are worked out from
dynamical background equations.

The linear evolution of density fluctuations is widely discussed in
BMM. The peculiar result, which is the starting point of this work, is
that coupled--DM fluctuations exhibit an almost $\beta$ indipendent
growth. In a synchronous gauge, it begins outside the horizon,
accelerating when fluctuations pass through it, and persisting
afterwards ($\tau \gg \tau_{hor}$), when the relativistic regime is
over. It is so all through the CI expansion period, as well as early
afterwards: in no case, after $\tau_{hor}$, coupled--DM fluctuations are
subject to stagflation. This occurs in spite of coupled--DM being a
fraction $\cal O$$(1\, \%)$ of the cosmic materials and while the
other components either freely stream, if uninteracting, or begin
sonic oscillations. The reason why this occur is recalled in the next
Section.

Then, by using the linear program discussed in BMM, we appreciate that
coupled--DM fluctuations, with amplitude $\cal O$$ (10^{-5})$ at
$\tau_{hor}$, reach an amplitude $ \sim 0.1$ at $\tau_{nl} \sim 500 \times
\tau_{hor}$. Non--linear effects start then to be significant. Afterwards,
at $\tau \sim 2$--$3 \times 10^3 \tau_{hor}$, it would then be $\langle
\delta_c \rangle \sim 1$, yielding $\delta \rho_c \gtrsim  \rho_c$;
linear results are then meaningless.

In this regime, we can achieve a reasonable insight on the actual
fluctuation evolution by following the behavior of (admittedly
unlikely) spherically symmetric density enhancements. Such approach,
in a different context, allows us to predict, e.g., the mass functions
of real physical systems, as galaxy clusters; it is so because the
approach yields a realistic {\it clock} of fluctuation growth and a
schematic picture of their~fate (see, e.g., \cite{PS}).

\begin{figure}[t!]
\begin{center}
\vskip -2.2truecm
\includegraphics[height=7.5cm,angle=0]{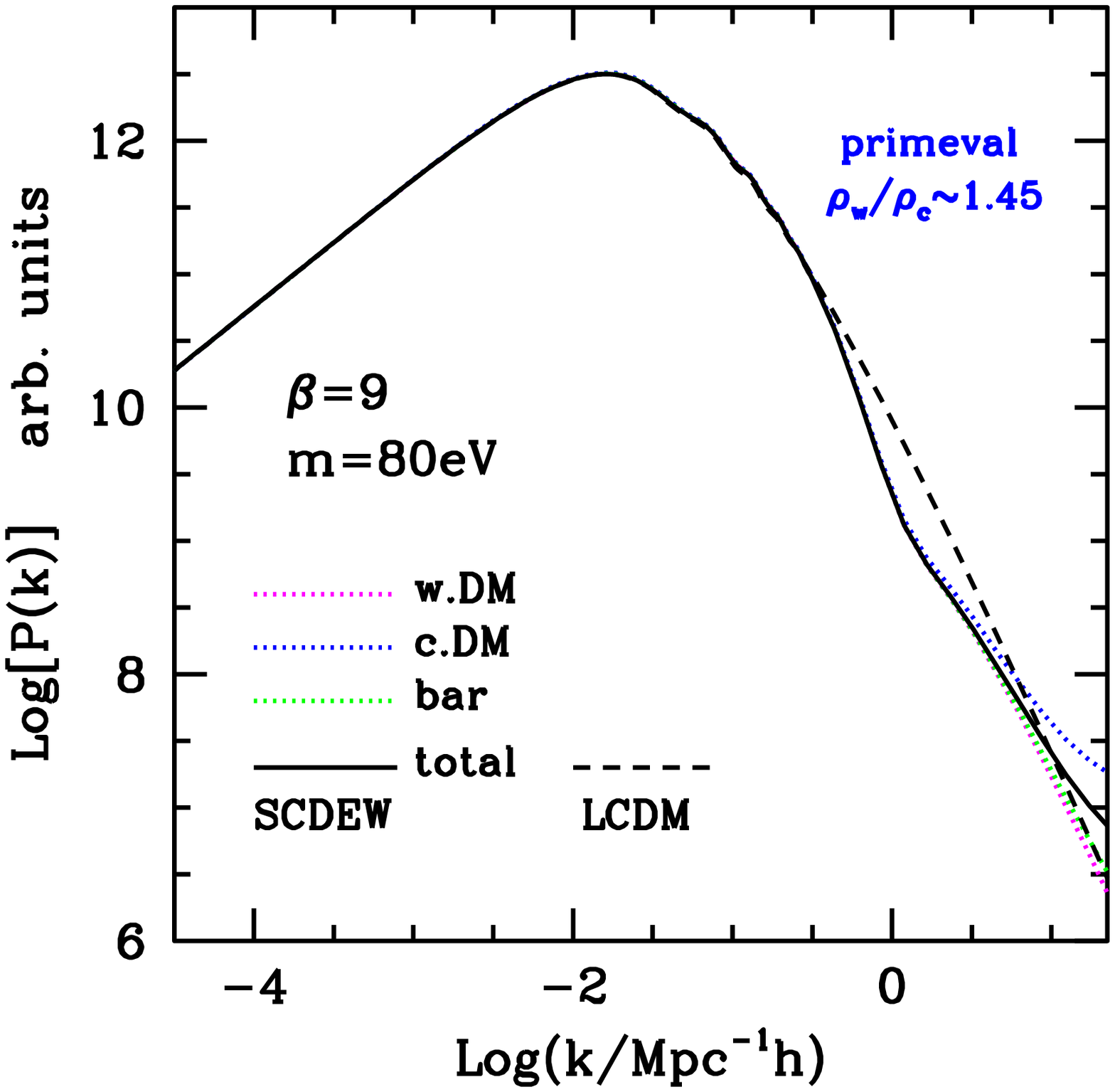}
\includegraphics[height=7.5cm,angle=0]{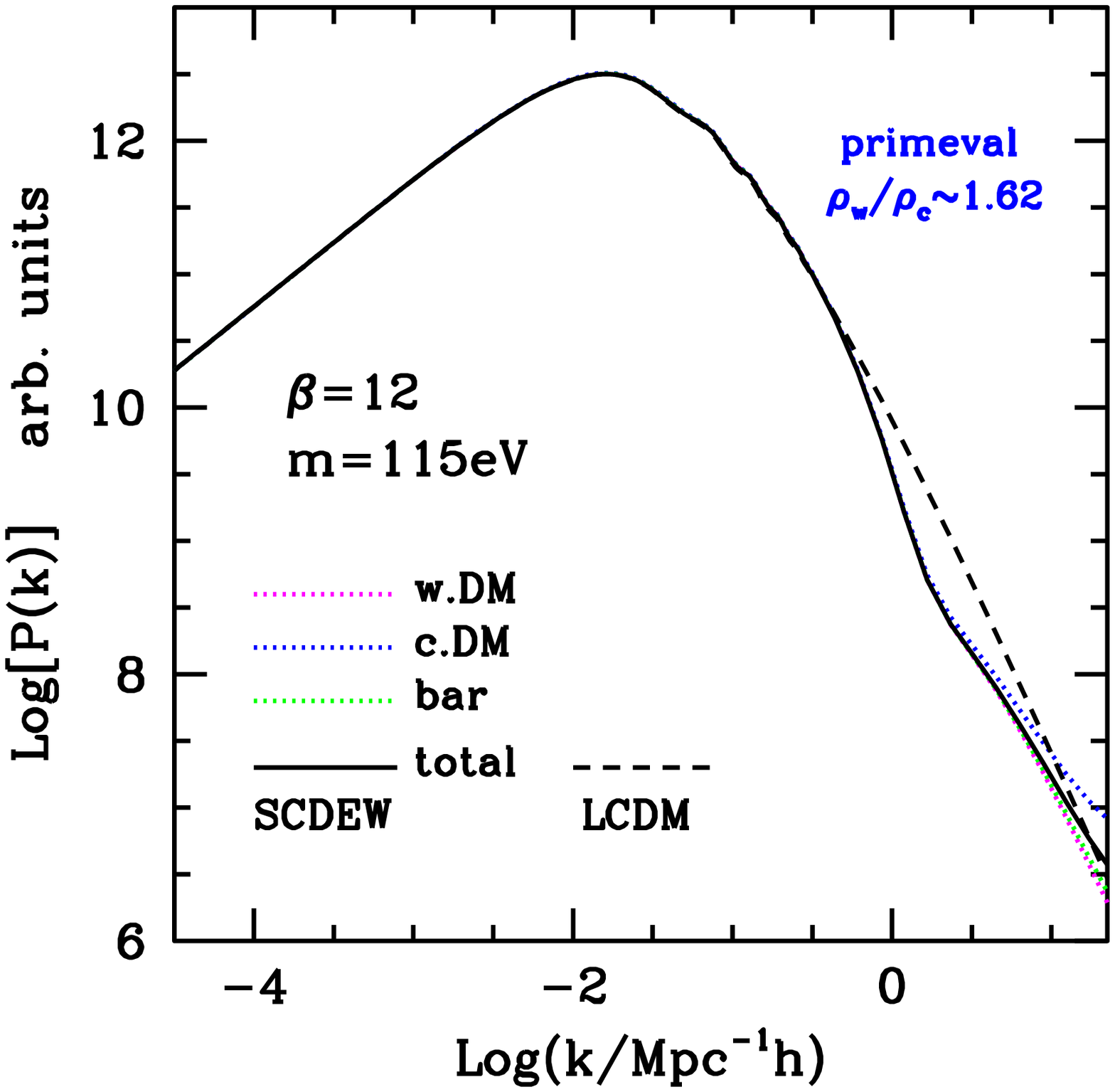}
\end{center}
\vskip -1.truecm
\caption{Linear spectra of the models used in this work, compared with
  LCDM spectra. The ratios between warm and coupled DM component
  densities, during the late CI expansion, is shown at the top right
  in blue. In both models we have a slight power deficit (up to a
  factor $\cal O$$(3)$) in the Lyman--$\alpha$ cloud region, while
  LCDM amplitude is recovered around average galactic scales. In
  principle, such increased amplitude means an earlier formation of
  structures on subgalactic scales. }
\label{spct}
\vskip +.1truecm
\end{figure}
However, at variance from what is done in the cited case, when we
start following the evolution of a fluctuation, here we shall not
assume it to expand within the Hubble flow, but work out its growth
rate from the linear regime. Results are then nearly independent from
the selected initial density contrast $\Delta_c = 1+\delta_c$ if
$\delta_c < 10^{-2}$. 

Most results of this paper are obtained by using 2 specific SCDEW
models: either $\beta = 9$ and $m \equiv m_w=m_c = 80\, $eV (model 9)
or $\beta = 12$ and $m \equiv m_w=m_c = 115\, $eV (model 12).  In
Figure \ref{spct} we show their $z=0$ linear spectra, as obtained from
our linear algorithm, discussed in BMM. In all models, at $z=0$,
$\Omega_b = 0.045$, $\Omega_\Phi = 0.70$, $h = 0.685$, with the usual
meaning of symbols. Furthermore we suppose $T_0 = 2.726\, $K and a
primeval helium abundance $Y_{He} = 0.24\, .$ Units yielding $c =
\hbar = 1$ are taken all through the paper.

The plan of the paper is as follows: In the next Section we derive the
equations needed to follow the density enhancement evolution until
virialization (phase I). In Section 3, we shall tentatively extend the
analytical treatment to the evolution after it (phase II).  In Section
4, numerical results will be shown. Section 4 contains a discussion of
the results found.

\section{Fluctuation evolution in the early Universe}
All through this paper, the expressions (\ref{meff1}) and
(\ref{beteff}) for DM particle mass and coupling will be used.
Accordingly, we shall never deepen in the fully CI regime. This is not
a problem, however, as the density enhancement behaviors, found for
the smallest $\tau_{hor}$ considered, are quantitatively identical to
those holding for $\tau_{hor}<\tau_{H}$. There exist, in fact, a long
period of {\it effective} CI regime, when CI violations are so small
to yield a negligible influence; see below for more details on this
point.

\subsection{A top--hat fluctuation in the early Universe}
Let us then consider a spherical top--hat overdensity, entering the
horizon with an amplitude $\delta_{c,hor}$, in the very early
Universe.
\begin{figure}[t!]
\begin{center}
\vskip -2.2truecm
\includegraphics[height=10.cm,angle=0]{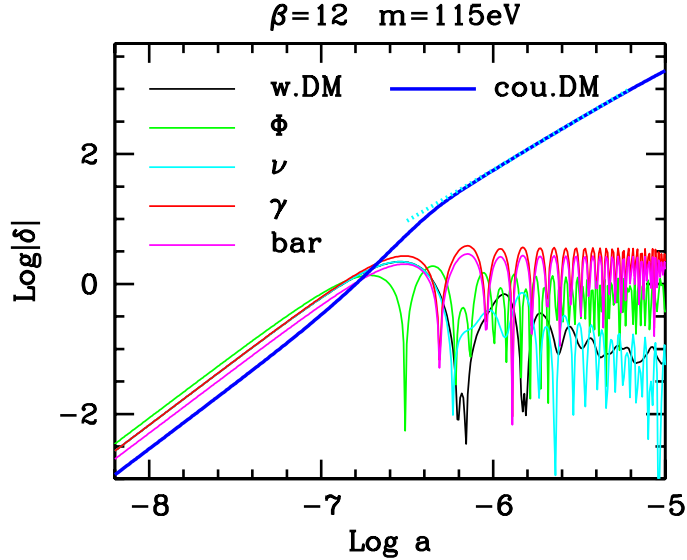}
\end{center}
\vskip -1.truecm
\caption{Typical linear evolution of linear density fluctuations close
  to horizon (the scale considered corresponds to $k=30\, h\,
  $Mpc$^{-1}$). The non--relativistic (linear) regime is approached
  when the cyan bended dotted line is attained (steepness $\delta_c
  \propto a^{1.6}$). The main parameters of the cosmology used are
  indicated at the top of the Figure, but most results are
  substantially model independent.  }
\label{flevo}
\vskip +.1truecm
\end{figure}
In this work we shall assume that $\delta_{c,hor} > 0$ and, mostly,
$\simeq 10^{-5}$, its top likelihood value. We expect fluctuations to
exhibit a Gaussian distribution, so that different (namely greater)
$\delta_{c,hor} $ values, although unlikely, are also possible. Our
treatment however holds only for $\delta_{c,hor}$ values small enough
to allow $\delta_c$ to enter a non--linear regime only when already
non--relativistic. The case of $\delta_c$ entering the non--linear
regime when still relativistic, in the frame of SCDEW models, relevant
for predictions on primeval Black Holes \cite{pBH}, will be discussed
elsewhere.

The critical point, however, is clearly illustrated in Figure
\ref{flevo}. Here we show the fluctuation growth, as derived from our
linear program, close to the horizon. At the horizon crossing, being
in the relativistic regime, coupled DM fluctuations exhibit a
significantly upgraded growth rate. When sufficiently inside the
horizon, the growth rate slows down, and $\delta_c \propto a^\alpha$
with $\alpha \simeq 1.6\, .$ The Figure also shows that coupled DM
fluctuations exhibit a steady growth while the other component
fluctuations undergo free streaming or enter a sonic regime with
stationary amplitude.

This behavior can be easily understood, at least in the
non--relativistic regime, by taking into account the treatment in
\cite{maccio2004}, concerning the evolution of coupled--DM
overdensities. {\bf Let us outline that this treatment, specifically
  devised to perform N--body simulations, holds both in the linear and
  in the non--linear regime}.  In that work, aiming to perform N--body
similations of coupled--DE models, it is shown that coupling effects
are equivalent to: (i) An increase of the effective gravitational push
acting between DM particles, for the density fraction exceeding
average, while any other gravitational action remains normal. The
increased gravitation occurs as though $G = 1/m_p^2$ becomes
\begin{equation}
G^* = \gamma\, G ~~~~{\rm with} ~~~~ \gamma = 1 + 4\beta^2/3
\label{gstar}
\end{equation}
($m_p:$ Planck mass).  (ii) As already outlined in eqs.~(\ref{meff1})
and (\ref{propto}), coupled--DM particle masses progressively
decline. This occurs while the second principle of dynamics still
requires that ${\bf f} = {\bf p}'$ (here the prime indicates
differentiation in respect to the ordinary time $t$). This yields the
dynamical equation
\begin{equation}
{d{\bf v} \over d\, t} = {{\bf f} \over m_{eff}} + \left|m_{eff}' \over
m_{eff} \right| {\bf v}~,
\label{fuma}
\end{equation}
i.e. an {\it extra--push} to particle velocities, adding to the
external force~{\bf f}. 

{\bf It should be outlined that, once eqs.~(\ref{gstar}) and
  (\ref{fuma}) are applied, the whole effects of coupling are taken
  into account; in particular, the (small) $\Phi$ field perturbations
  cause no effect appreciable at the Newtonian level (see again
  \cite{maccio2004}). This is true even in the presence of extreme DM
  density contrasts, as those found in the halos produced by N--body
  simulations, and, even more, in the linear case considered here.} 

The self--gravitational push due to $\delta_c$ is then $\propto G^*
\rho_c \delta_c = G \rho_{cr} (2/3) \delta_c \times (1 + 3/4\beta^2)$,
with the last factor exceeding unity just by $\sim 1\, \%$ for $\beta
= \cal O$$(10)$. Henceforth, coupled DM fluctuations, in the
non--relativistic regime, grow as though concerning the total cosmic
density $\rho_{cr}$, at least. The slightly reduced amplitude, in
fact, is overcompensated by the {\it extra--push} and, as previously
outlined, the linear program gives evidence of a grows $\propto
a^\alpha$ with $\alpha \sim 1.6\, .$

Within this context, we can schematically describe the evolution of a
spherical top--hat density enhancement of amplitude $\delta_c$. The
fluctuation initially expands according to linear equations, but, as
soon as $\delta_c \gtrsim 10^{-2}$--$10^{-1}$, non--linear effects
become no longer negligible. At this stage, the radius $R$ of the
top--hat, growing more slowly than the scale factor $a$, reaches a
maximum value $R_{top}$ and then starts to decrease.  Eventually,
however, inner kinetic and potential energies reach a virial balance,
when the sphere has a radius $R_{vir}$.

Let us then recall that, in the framework of a ``standard CDM''
cosmology, there exist an analytical (parametric) solution of the
equations ruling the evolution of a spherical top--hat overdensity.
The evolution of a spherical overdensity in a coupled--DE model, with
$\beta \ll 1\, ,$ was considered by \cite{Msolo}. The key issue was
then that both baryons and coupled DM fluctuations grow, but at
different rates: the DM component is faster in reaching maximum
expansion and starts to recontract before baryons. It is then
necessary to share the top--hat fluctuation into shells, which
gradually compenetrate. The number of shells needed is determined by
the precision wanted. In Figure \ref{flevo} we directly see that we
are now dealing with a case when only coupled--DM fluctuations
grow. The equations ruling $R$ evolution are then similar to those
obtained in \cite{Msolo}, with the welcomed difference that we need no
subdivision into spherical shells.

The relation between the comoving sphere radius $c = R/a$ and the
density contrast $\Delta_c =1 + \delta_c$ then reads
\begin{equation}
\Delta_c = 1+\delta_c = \Delta_{c,r} c_r^3/c^3~,
\end{equation}
as the subscript $_r$ refers to a suitable reference time;
accordingly, by assuming $\delta_c \propto \tau^\alpha$,
\begin{equation}
{\dot c \over c_r} = -{\alpha \over 3} {\delta_{c,r} \over \Delta_{c,r}}
{1 \over \tau}
\label{dotc}~;
\end{equation}
this relation allows us to chose arbitrarily the time $\bar \tau$,
during the linear regime, when we start to use $c$ instead of
$\delta_c$ to follow the top--hat dynamics.

\subsection{Dynamical equation}
In strict analogy with eq.~(9) in \cite{Msolo}, the evolution of the
overdensity then follows the equation
\begin{equation}
\ddot c = -\left( {\dot a \over a} - C\dot \Phi \right) \dot c  -\gamma G {1
  \over ac^2} [M(<R) - \langle M(<R) \rangle]~.
\label{eq9}
\end{equation}
Here, as in eq.~(\ref{dotc}), derivatives are taken in respect to the
conformal time $\tau$; $\Phi$ is the background value of the scalar
field. Furthermore, $M(<R)$ is the actual mass within $R$, while
$\langle M(<R) \rangle$ is the {\it average} mass in a sphere of
radius $R$, but, if assuming all components but coupled--DM to be
unperturbed, we only need evaluating
\begin{equation}
G \langle M_c(<R) \rangle = G {4\pi \over 3} \rho_{cr} \Omega_c a^3 c^3
= {\Omega_c \over 2} {h_2 \over \tau^2} a\, c^3
\label{GbarM}
\end{equation}
with
\begin{equation}
h_2 = {8 \pi \over 3} G \rho_{cr} a^2 \tau^2
\label{h2}
\end{equation}
being close to unity, during the effective CI expansion {and
  exactly unity at $\tau < \tau_H$, when also $\Omega_c \equiv
  1/2\beta^2$.
However, at later times, when $\beta$ must be replaced by
$\beta_{eff}$, no similar relation holds.

Let then $\bar \Delta$ be the density contrast at the time $\bar \tau$,
so that
\begin{equation}
{1 \over \bar \Delta} G M_c(<R) = G {4\pi \over 3} m_{eff}(\tau)\, \bar
n_c\, \bar a^3 \bar c^3~.
\label{1od}
\end{equation}
Here $n_c$ is the number density of coupled--DM particles, whose mass
$m_{eff}$ is given by eq.~(\ref{meff1}) (all ``barred'' quantities
refer to the ``initial'' time $\bar \tau$). {As $n_c a^3$ is
  constant in time, it is also
\begin{equation}
{1 \over \bar \Delta} G M_c(<R) = 
 G {4\pi \over 3} m_{eff}\, n_c ~a^3 \bar c^3 = 
 G {4\pi \over 3} \rho_{cr} \Omega_c ~a^3 \bar c^3 = 
{\Omega_c \over 2}{h_2 \over \tau^2} a\, \bar c^3 ~.
\label{seconda}
\end{equation}
Accordingly, by setting $\Delta M_c = [M_c(<R) - \langle M_c(<R) \rangle]
$, we have
\begin{equation}
 G ~\Delta M_c = {\Omega_c \over 2} {h_2 \over
   \tau^2} a \bar c^3 \left({\bar \Delta}-x^3 \right) ~~~{\rm with} ~~~
 x=c/\bar c
\label{2ter}
\end{equation}
so that
\begin{equation}
 {G \over ac^2} [M(<R) - \langle M(<R) \rangle] = {\Omega_c \over 2}
 {h_2 \over \tau^2 x^2} \bar c \left({\bar \Delta}-x^3 \right)~.
\label{2term}
\end{equation}
In turn,} the difference $h_0 = \dot a/a-C\dot\Phi$ exactly vanishes,
during the early CI expansion, both terms being then $1/\tau$. As the
background density of coupled DM fulfills the equation
\begin{equation}
\dot \rho_c + 3 (\dot a/ a) \rho_c = -C \rho_c \dot \Phi
\label{dotrhoc}
\end{equation}
it is however worth keeping into account that
\begin{equation}
h_0 = -C\dot \Phi + {\dot a \over a} = {\dot \rho_c \over \rho_c} +
4\, {\dot a \over a}~,
\label{f0}
\end{equation}
as this allows an easier numerical evaluation of $h_0$. Altogether,
eq.~(\ref{eq9}) also reads
\begin{equation}
\ddot x = -h_0 \dot x - h_1
\left(\bar \Delta - x^3 \right) {1 \over u^2 x^2}
\label{new9}
\end{equation}
with 
\begin{equation}
h_1 =  {1 \over 2} \gamma \Omega_c h_2~,~~~
\label{PQ}
\end{equation}
and $u = \tau /\bar \tau$, while, at variance from elsewhere, {\it
  dots here indicate differentiation in respect to $u$}. In
eq.~(\ref{new9}), describing a process due to self--gravity, the
gravitational constant no longer explicitly appears, being reabsorbed
in the definition of $h_2$ (eq.~\ref{h2}) and then in~$h_1$.

Let us finally outline that, until we are close to the CI expansion,
the coefficient
\begin{equation}
{1 \over 2} \gamma \Omega_c = \left(1 + {4 \over 3} \beta^2 \right)
{1 \over 4 \beta^2} = {1 \over 3} + {1 \over 4 \beta^2}
\label{gammaomega2}
\end{equation}
keeps close to 1/3, for reasonable $\beta $'s (see however Figure
\ref{gom}). Eq.~(\ref{new9}) however holds both then (when also
$h_2=1$) and when CI is abandoned, so that $h_1$ can become even quite
different from 1/3~.
\begin{figure}[t!]
\begin{center}
\vskip -5.truecm
\includegraphics[height=10.cm,angle=0]{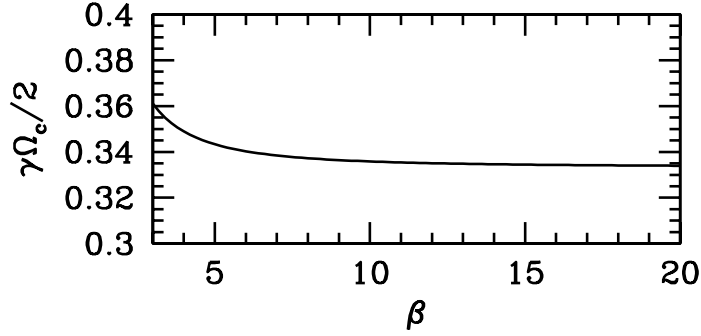}
\end{center}
\vskip -.8truecm
\caption{During the CI expansion the coefficient $h_1 = {1 \over
    2}\gamma \Omega_c$ is almost $\beta$ independent. }
\label{gom}
\end{figure}

\subsection{Virialization}
Numerical solutions of eq.~(\ref{new9}) yield the expected growth and
successive recontraction of the radius $R$ of top--hat density
enhancements,  as well as the gradual increase of the density
  contrast $\Delta_c$ in respect to the average coupled--DM density
  $\rho_c$; it is also clear that an {\it ideal} top--hat would
expand and recontract, according to the above expressions, down to a
relativistic regime. Top--hat fluctuations were however considered
because their equations of motions are integrable and provide an
insight into the real timing of true fluctuation evolution.
Virialization is then the successive step assumed to occur, because
real motions are unordered and exact sphericity breaks down when we
pass from expansion to recontraction.

 To establish the conditions for virial balance, we then need the
  expressions of the kinetic and potential energy for the sphere. In
accordance with \cite{Msolo}, the kinetic energy expression is rather
simple to obtain and reads
\begin{equation}
T_c(R) = {3 \over 10} M_c {R'}^{\, \, 2}~,
\label{Tc}
\end{equation}
 as the factor 3/10 derives from
integration on a sphere. Here, the prime indicates differentiation with
respect to ordinary time, so that $ R' = (a\, \dot c + \dot a\, c )/a
= \dot c + (\dot a/a) \, c $, if dots indicate differentiation with
respect to $\tau$; by using eq.~(\ref{h2}), we then have
\begin{equation}
2 \times {5 \over 3} {T_c \over M_c} = {\bar c^2 \over \bar \tau^2} 
\left(\dot x + {h_2^{1/2} \over u} x \right)^2~,
\label{kinetic}
\end{equation}
dots indicating here differentiation in respect to $u$.

The potential energy is then made of two terms, arising from DM
fluctuation interacting with DM background and all backgrounds
interacting with themselves. Therefore, in agreement with \cite{Msolo}
where, however, the only unperturbed background was DE,
\begin{equation}
{U_c(R) \over M_c} = -{3 \over 5} G {[\langle M_c \rangle + \gamma
    \Delta M_c] \over R } -{4 \pi \over 5}G \rho_{back} R^2 = -{3
  \over 5}  \gamma G { \Delta M_c \over R } -{4 \pi \over 5}G
  \rho_{cr} R^2~.
\label{Ur}
\end{equation}
By using the expression (\ref{2ter}) for $G\, \Delta M_c$, we then have
$$
 - {\gamma G \over R} \Delta M_c = - {\gamma \Omega_c \over 2}
{h_2 \over \tau^2 x} \bar c^2 (\bar \Delta - x^3) = -
{\bar c^2  \over x}{h_1 \over \tau^2}
 (\bar \Delta - x^3)
$$
while
$$
-{4 \pi \over 3} G\rho_{cr} R^2 = 
-{8 \pi \over 3} G\rho_{cr} a^2 {c^2 \over 2} = 
-{\bar c^2 \over \tau^2} {h_2 x^2 \over 2}
$$
so that
\begin{equation}
{5 \over 3} {U_c(R) \over M_c} = - { \bar c^2 \over \tau^2} \left[{ h_2
    x^2 \over 2} + {h_1 \over x} (\bar \Delta -x^3 ) \right]
\label{potentia}
\end{equation}
Virilization is then obtainable by requiring that
\begin{equation}
2 \times {5 \over 3} {T_c \over M_c} +
{5 \over 3} {U_c \over M_c} = 0
\label{virial}
\end{equation}
or, by using the expression (\ref{kinetic}) and (\ref{potentia})
hereabove, 
\begin{equation}
(u \dot x + {h_2^{1/2}} x)^2 - h_1 (\bar \Delta/x
-x^2 ) - h_2 x^2/2 = 0~.
\label{virial1}
\end{equation}
From the $c_v$ and $\tau_v$ values fulfilling this equation, we then
derive tha virial radius $R_v = c_v a_v$. Neither the equation of
motion, nor this expression, are suitable for analytical treatment.

\section{After virialization}
In order to better understand the physical sense of the virialization
condition, let us assume that the CI expansion regime holds and, in
particular, $h_0=0$ and $h_2=1$. According to eqs.~(\ref{Tc}) and
(\ref{Ur}), the top--hat virial then reads
\begin{equation}
{5 \over 3} {Vir \over M_c} =  \langle v^2 \rangle - {4\pi \over 3} G
\rho_{cr} \left( \gamma \Omega_c \Delta + {1 \over 3} - \Omega_c
\right)R^2~,
\label{wireq}
\end{equation}
once we replace ${R'}^2 = \langle v^2 \rangle$, as we expect particle
velocities yielding coherent contraction to turn into randomly
distributed speeds.

It is then convenient to multiply this relation by $m_{eff}^2$ and
outline the vanishing of the virial through the approximate relations
\begin{equation}
\langle p^2 \rangle = \gamma G {N_c m_{eff}^3 \over R_v} =
{4\pi \over 3} \gamma G \rho_{cr,v} \Omega_c
\Delta_v R_v^2 m_{eff}^2 = {1 \over 4t_v^2} h_1 \Delta_v R_v^2
m_{eff}^2 ~,
\label{wireq1}
\end{equation}
($N_c$ is the total number of coupled--DM particles, yielding a total
mass $N_cm_{eff}$) so to take easily into account that, in spite of
$m_{eff}$ progressive decrease, the square averaged momentum $p_v^2
\equiv \langle p^2 \rangle$ is however expected to keep constant. All
quantities with index $_v$ refer to virialization. The relation
(\ref{wireq1}) is readily understandable as a balance between twice
the average particle kinetic energy $\langle (p^2/m_{eff}) \rangle$
and its potential energy in respect to coupled--DM (only), just as in
the process of virialization of a top--hat matter density enhancement
after matter--radiation decoupling (PS--case).

In both cases, once $R_v$ is reached, the recontraction process is not
immediately discontinued and a stationary configuration is reached
after a few oscillations; since $t_v$, however, the average particle
momentum is expected to keep $p_v$. The point is then that the average
particle momentum, at any $t>t_v$, exceeds the virial equilibrium
momentum
\begin{equation}
p_v^2(t) = p_v^2(\tau_v/\tau)^3 = p_v^2(t_v/t)^{3/2}~,
\label{latmom}
\end{equation}
because of the progressive decrease of $m_{eff} \propto \tau^{-1}$.
Should all particle momenta coincide with $p_v$, a global free
streaming would follow, and the characteristic time for a full
dissolution would be the crossing time $t_{cross} = R_v/v_v$ (here
$m_{eff}v_v=p_v$). The distribution of particle momenta, however,
can be expected to be close to a maxwellian
\begin{equation}
f(p/p_t) \equiv f(x) = {4 \over \sqrt{\pi}} x^2 e^{-x^2}~,
\label{boltzmann}
\end{equation}
with $p_t = \sqrt{2/3}\, p_v$ being the distribution top. Evaporation
will then start from fastest particles and, in principle, it is
possible that the momentum they carry away allows a sufficient average
momentum decrease, so that 
\begin{equation}
\langle p^2(t) \rangle = p_v^2 {N_c(\tau) \over \bar N_c} \left(\tau_v
\over \tau \right)^3 = p_v^2 {N_c(t) \over \bar N_c} \left(t_v
\over t \right)^{3/2}~.
\label{neweq}
\end{equation}  
Let us outline that $\langle p^2(t) \rangle$ is required to decrease
faster than $p_v^2(t)$, as the potential energy to balance also
decreases when the total number of particles ($N_c$) declines.  By
dividing both sides of eq.~(\ref{neweq}) by $p_v^2$ and taking into
account the particle distribution, we then have
\begin{equation}
{p_t^2 \over p_v^2}  {\int_0^\alpha dx \, x^2 f(x) \over
\int_0^\alpha dx f(x)} = 
 {2 \over 3} {\int_0^\alpha dx\, x^2 f(x) \over
\int_0^\alpha dx \, f(x)} = 
\left(1 \over 1+ t_{\alpha}/t_v  \right)^{3/2}
\int_0^\alpha dx \, f(x)~;
\label{eq3}
\end{equation}
here we took into account that the Boltxmann distribution is
normalized to unity; we also set $t = t_v+t_{\alpha}$, so to
outline the time $t_{\alpha}$ when particles with momenta $p >
\alpha\, p_t$ were allowed to evaporate. We can also define
\begin{equation}
F(\alpha) \equiv {2 \over 3} {\int_0^\alpha dx {4 \over \sqrt{\pi}}
  x^4 e^{-x^2} \over \left[\int_0^\alpha dx {4 \over \sqrt{\pi}} x^2
    e^{-x^2}\right]^2} = \left(1 \over 1+ t_{\alpha}/t_v \right)^{3/2}
\label{eq4}
\end{equation}
and seek the value $\alpha_m$ minimizing $F(\alpha)$; the point is that,
after the most rapid particles have evaporated, the momentum decrease
granted by further slower particle evaporation is beaten by the
potential energy decrease due to the outflow of particles belonging to
the bulk of the distribution. 

The process should then not proceed beyond $t_{ev} \equiv
t_{\alpha_m}$, that we dubb {\it evaporation time}; i.e., in order to
grant a ``long life'' to the residual density enhancement, it should
be $ t_{cross} \ll t_{ev}$. This would allow only particles with high
momentum to outflow, even though initially well inside the
enhancement. Furthermore, $t_{cross}$ is also the order of magnitude
of the time needed for rearranging particle momentum distribution,
when the fastest particles have outflown, so to recover a Bolzmann
distribution, and allow for further fast particle evaporation.

According to eq.~(\ref{wireq1}), it is then easy to see that
\begin{equation}
\label{tcross}
{t_{cross} \over t_v} = {2 \over (h_1 \Delta_v)^{1/2}}~.
\end{equation}
Notice that eq.~(\ref{tcross}) holds also in the PS--case provided we
replace $h_1$ by $\Omega_m/2$ (matter density parameter). A fair
comparison between evaporation and crossing times then requires 
that $\Delta_v$ is known.

Before passing to a numerical treatment of the problem, let us however
outline that the point we still debate is the time--scale for the
top--hat dissolution, which is however expected to occur anyhow.
Should however be $t_{ev} < t_{cross}$, we face a situation when the
dissolution is almost immediate, taking just a few $t_{cross}$; i.e.,
the time needed to settle in virial equilibrium, in the PS--case.

\section{Numerical treatment}

\begin{figure}[t!]
\begin{center}
\vskip -1.truecm
\includegraphics[height=11.32truecm,angle=0]{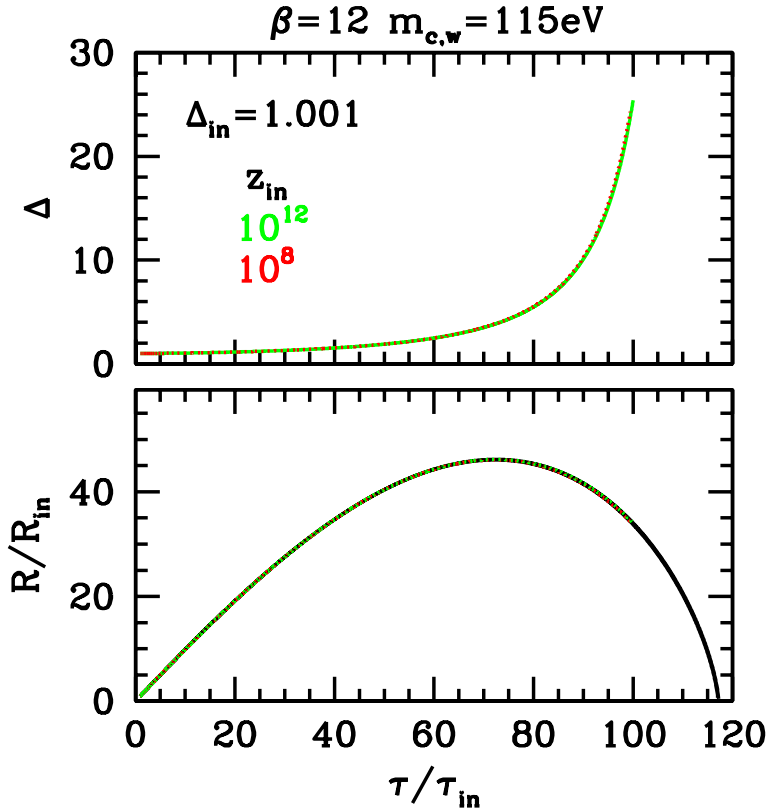}
\vskip -.3truecm
\caption{Top hat radius and density contrast evolution for the model
  12. In abscissa, the conformal time $\tau/\tau_{in}$ is
  substantially coincident with $a/a_{in}$. Both $z_{in} = 10^{12}$
  and $10^8$ are considered, but the curves essentially overlap. The
  only tenuous difference is a slight shift for the value of the final
  density contrast. In the bottom plot, the eventual evolution of $R$,
  assuming virialization not to occur, is also shown down to $R=0$
  (relativistic regime).  }
\label{XRvsa12}
\vskip +.1truecm
\end{center}
\end{figure}

\subsection{Phase I}
Top--hat evolution, during the CI expansion, is fairly easily
integrated, as the dynamical coefficients are then constant. We shall
not report the results for this case, but only those obtained by
setting the initial condition at $z_{in} = 10^{12}$ and $10^8$. As a
matter of fact, the former case yield results numerically coincident
with those for $z_{in} > z_H$, while the very difference between $z_{in}
= 10^{12}$ and $10^8$ is quite small.

In Figure \ref{XRvsa12} we show the behaviors of the radius $R$ and
the density contrast $\Delta$ {\it vs.} the conformal time $\tau$
for the model 12.
\begin{figure}[t!]
\begin{center}
\vskip -.4truecm
\includegraphics[height=11.32truecm,angle=0]{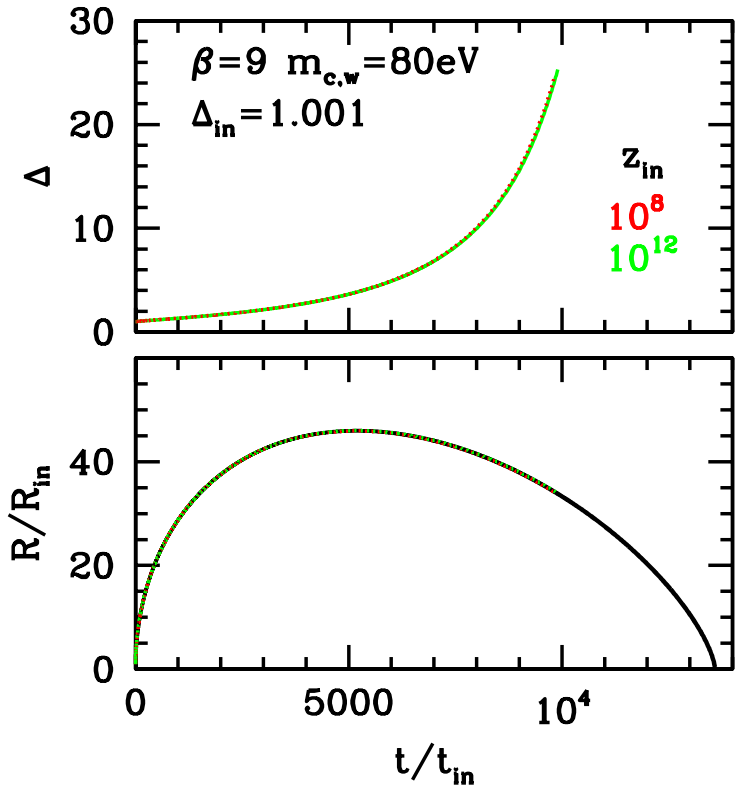}
\vskip -.3truecm
\caption{Top hat radius and density contrast evolution for the model
  9. In abscissa, the ordinary time $t/t_{in}$. Also here the curves
  for $z_{in} = 10^{12}$ and $10^8$ essentially overlap while, again,
  the only tenuous difference is a slight shift for the value of the
  final density contrast. In the bottom plot, as in the previous
  Figure, the eventual evolution of $R$, assuming virialization not to
  occur, is also shown down to $R=0$.  }
\label{XRDvst.9-3}
\vskip +.1truecm
\end{center}
\end{figure}
For model 9 we then rather show the evolution by using ordinary time
in abscissa (figure \ref{XRDvst.9-3}).

As a matter of fact, it seems more significant to outline the
different apparent behavior when the abscissa is changed, rather than
the tiny model dependence.

\begin{figure}[t!]
\begin{center}
\vskip -.4truecm
\includegraphics[height=12.32truecm,angle=0]{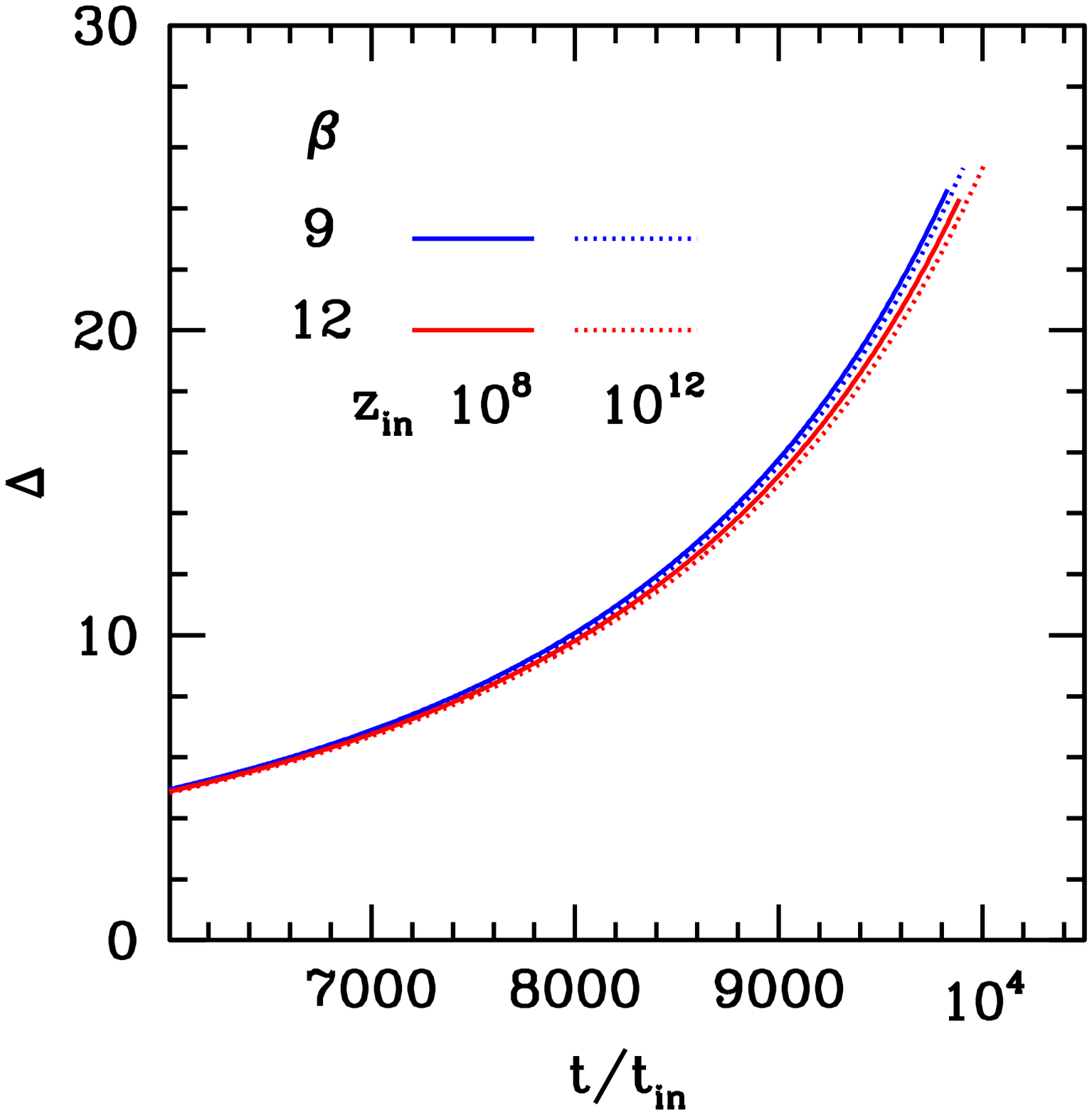}
\vskip -.3truecm
\caption{Blow up of the final part of the density contrast growth,
  before virialization, for both models and $z_{in}$.  }
\label{XRDa2-3}.
\vskip -2.9truecm
\end{center}
\end{figure}
In order to magnify the differences between models and $z_{in}$
values, where they exist, we however plot the final part of the
density contrast increase, prior to virialization, in Figure
\ref{XRDa2-3}.  For the sake of completeness let us then provide the
numerical values of the virial density contrasts. For $\beta=12$ they
are 25.4 or 24.3 when $z_{in} = 10^{12}$ or $10^8$, respectively,
while, for $\beta=9$, the corresponding values are 25.3 or 24.6~. Let
us also add that, if we start following the density contrast evolution
when $\bar \Delta =1.1$, instead of 1.001, we obtain slightly smaller
values: e.g., for $\beta=12$, they read 25.3 and 24.2, if the density
contrast $\Delta_c=1.1$ is attained at $z_{in} = 10^{12}$ or $10^8$,
respectively. Non--linearity, when $\delta_c < 0.1$, has quite a
limited impact, not exceeding half percent.

This greater ``initial'' density contrast can be due to rare
fluctuations entering the horizon already with a wider amplitude; the
point is that we however reach a virial density contrast just
marginally different from starting when $\Delta_c=1.001$; i.e., that
non linearity effects are negligible for $\delta_c < 0.1$. The
procedure followed is therefore well approximated also if the
relativistic regime due to horizon crossing ends up when the density
contrast is already $\cal O$$(1.1)$.  Such greater $\Delta_c$ at
$z_{in}$, however, is not met just for exceptional fluctuations, being
possibly due just to an earlier horizon crossing and, therefore, to
fluctuations over smaller scales.

Let us rather outline that a final density contrast $\sim 25$--26 is
``small''. As the fractional contribute of coupled DM to the overall
density is $\Omega_c \sim 0.005$, it is clear that even $\Delta_v
\Omega_c \sim 0.13$ is still far from unity: the total density
enhancement, at virialization, does not exceed the overall density, as
though still being in a {\it quasi}--linear regime. In turn, this
strengthens the reliability of results obtained by neglecting density
fluctuations in other components, when considering coupled--DM
fluctuation evolution. Taking them into account could only yield a
modest correction for final results.

It can also be significant to follow the evolution of the dynamical
coefficient $h_i$. In Figure \ref{Xparam} we show them for the case
$\beta=12$.
\begin{figure}[t!]
\begin{center}
\vskip -.4truecm
\includegraphics[height=10.32truecm,angle=0]{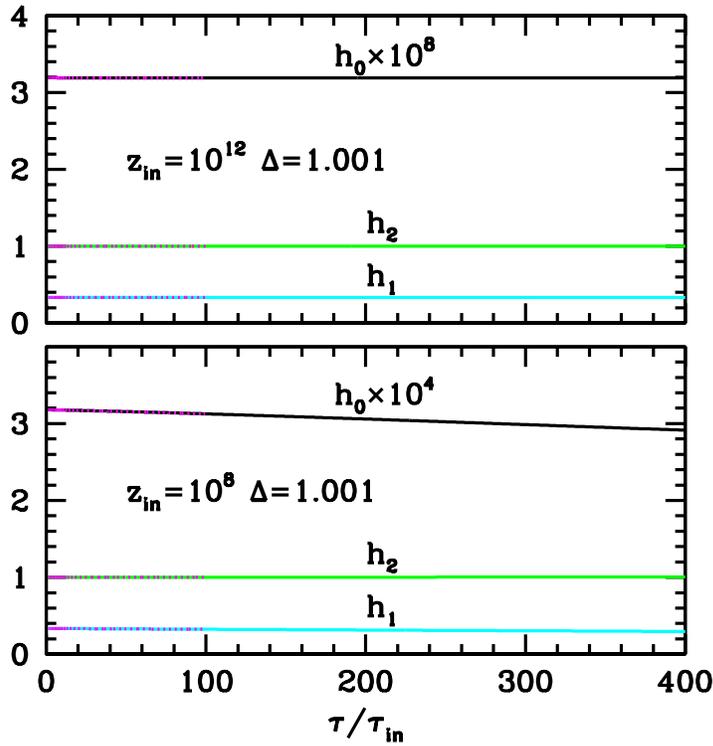}
\vskip -.3truecm
\caption{Dynamical coefficients for the model 12. At the l.h.s., the
  $\tau$ interval covered before virialization is dotted in magenta
  color. Notice also: (i) When $z_{in}=10^{12}$ all coefficient appear
  substantially constant, even behind the l.h.s. interval; on the
  contrary, for $z_{in}=10^8$ a slight time dependence is appreciable,
  namely for $h_0$. (ii) The difference of $h_0$ from nil rapidly
  increases when smaller $z_{in}$ values are considered; however, even
  $h_0 \sim 10^4$ values yield no contribution to the actual $R$
  evolution. }
\label{Xparam}.
\vskip -2.9truecm
\end{center}
\end{figure}
The behavior for $\beta=9$ does not exhibit significant differences.

In the Figure, the ranges of $h_i$ values used by the numerical
integrator, before virialization, are outlined by magenta dots.
Namely on such interval, $h_i$ variations are however quite small.

Notice however how $h_0$ increases when smaller $z_{in}$ values are
considered; during the CI expansion, $h_0 \equiv 0$; although so
small, the values of $h_0 \neq 0$ shown outline the exit from CI
expansion. In spite of that, even $h_0 \sim 10^4$ values yield no
appreciable contribution to the numerical $R$ evolution.

Notice also that $h_i$ coefficient appear fully $\tau$ independent,
for $z_{in}=10^{12}$; on the contrary, for $z_{in}=10^8$ a time
dependence is appreciable; it is strongest for $h_0$, but, as earlier
outlined, this bears no appreciable dynamical consequences.

\subsection{Phase II}
Once the density contrast $\Delta_v$ is known, the crossing time
(\ref{tcross}) can be soon evaluated, being $t_{cross} \simeq 0.7\,
t_v$ (in the PS case, for $\Lambda$CDM, $t_{cross}\simeq 0.5\, t_v $).

In Figure \ref{evospect} we report the $F(\alpha)$ dependence,
showing that $F(\alpha_m) \simeq 0.81$, so yielding
\begin{figure}[t!]
\begin{center}
\vskip -.7truecm
\includegraphics[height=7.cm,angle=0]{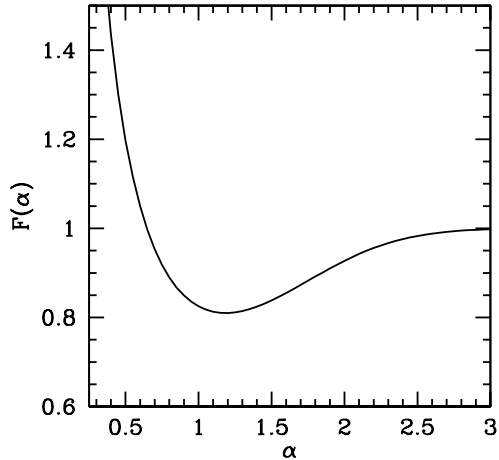}
\end{center}
\vskip -.5truecm
\caption{The dependence on $\alpha$ of the $F(\alpha)$ allows us to
  estimate down to which momentum ($\alpha \times p_t$) it is
  dynamically ``convenient'' that fast particles evaporate from the
  density enhancement, to ease a virial equilibrium recovery, in spite
  of $m_{eff}\propto \tau^{-1}$. The fast $F$ increase, when $\alpha$
  shifts below $\sim 1.21$, arises because the bulk of the Boltzmann
  distribution is then supposed to outflow.  }
\label{evospect}
\end{figure}
$t_{ev} = 0.15\, t_v$. The conclusion is that, once the virial
equilibrium condition is attained, the density enhancement is unable
to settle on it. The expected downward and upward oscillations which,
in the PS case, last $\sim t_v$, here are slightly longer and doomed
to end up with a substantial particle free streaming.

A way to stabilize the virialized system can only exist if, during the
fluctuation linear and/or non--linear growth, other cosmic component
particles were allowed to accrete, as will be when the WDM component
approaches derelativization.

\section{Discussion}
In a standard cosmological model with warm DM made of particles with
mass $\sim 100\, $eV, the minimal fluctuation scale surviving free
streaming is the scale entering the horizon at $z_{der} \sim 6 \times
10^{\, 5} (m_w/100\, {\rm eV})^{4/3} (\Omega_w h^2)^{-1/3}$, so
ranging about $2 \times 10^{13} h^{-2} M_\odot$ and exceeding the size
of the largest galaxies. The presence of coupled--DM in SCDEW models
allows us to shift the critical redshift from $z_{der}$ to $\sim 10^4
z_{der}$, so lowering by $\sim 12$--13 orders of magnitude the mass
scale of the minimal surviving WDM fluctuation.

The peculiarity of SCDEW cosmologies, however, is that this is not due
to an {\it ad--hoc} mechanism, being the unavoidable consequence of
the previous expansion along an attractor, through modified radiative
eras. SCDEW cosmologies, infact, are characterized by a substantial
modification of such early expansion regime, i.e., the constant
presence of coupled DM and $\Phi$, in fixed proportions, aside of
standard radiative components.

The above result is obtained by studying the evolution of a top--hat
density fluctuation in the late radiative era. Using a spherical
fluctuation to work out the expected time scale of processes is not a
new procedure. In a different context, it was first applied to predict
the mass function of cosmic bounded structures, as galaxy clusters
\cite{PS}.  Results were excellent and, with suitable improvements, a
similar approach is still in use.

When we treat a top--hat density enhancement of radius $R$, we find
$R$ gradually slowing down its growth rate in respect to the scale
factor $a$. Eventually, the $R$ increase stops and $R$ begins to
decrease. After a suitable time, however, kinetic and potential energy
reach a virial balance, so that we should expect equilibrium to be
attained.

Here however comes the most peculiar feature of coupled--DM
fluctuations: the virial condition is unstable. This is due to the
progressive decrease of the coupled--DM particle mass $m_{eff}$,
causing the increase of the kinetic energy $\sim p^2/m_{eff}$ if the
momentum $p$ is conserved, and a symultaneous decrease of the depth of
the potential well, roughly $\propto m_{eff}^2$.

As a consequence of these variations, the most rapid particles
  are expected to evaporate. We provided analytical tools to estimate
  evaporation effects and, as above outlined, estimated how long a
  significant density contrast can persist after virialization.

It is however legitimate to wonder how reliable can be estimates based
on a spherical top--hat evolution. The physics described here,
however, does not seem to need a sphericity assumption. Quite in
general, coupled--DM particles, embedded in a fluctuation entering the
horizon, are initially slow enough, so that their kinetic energy does
not interefere with fluctuation growth. The evolution described by
linear programs occurs under such conditions, but eventually causes a
fast growth of coupled--DM fluctuations, in spite of their density
being a small fraction of the total density. The reach of the
non--linear regime, therefore, is independent from any spherical
modeling.

We then expect that non--linearity produces significant energy jumps,
and the possibility to transfer such potential energy jumps onto
particle kinetic energy. Once particle momenta reach a significant
level, particle velocities burst, aside of $m_{eff}$ decrease. Escape
velocity could then be approached and overcame, and the heaten up
coupled--DM could no longer be constrained in primeval
inhomogeneities. The study of top--hat spherical fluctuations tries to
model these events and, hopefully, to provide a reasonably reliable
clock for them.

\end{document}